# DESIGN OF AN ALGORITHM FOR ACOUSTIC SIGNAL DETECTION OF MOVING VEHICLES

Daniel Blasco Avellaneda*[1]

*blasco@ugr.es
[1]Applied Physics Department, University of Granada

**ABSTRACT**

The precise detection and counting of vehicles is fundamental to know the state of agglomeration of the roads, as well as to anticipate possible actions of regulation of traffic or future changes that it is necessary to carry out in the urban design in the management of the transport. It is in fact part of the programs of many municipalities for statistical purposes, for the planning, design, extension and modeling of structures.

The operator L.O.G. (Laplacian of gaussian) is commonly used for the detection of elements in images, by the analysis of its border, and is a widely extended method in image filtering. However, as such is not an extended procedure in the analysis of one-dimensional signals.

We developed a procedure inspired by LOG filtering, and performed the analysis of the acoustic signal obtained by a portable recorder located on the side of a single lane road, by evaluating the negative minima of the second discrete derivative of the smoothed acoustic signal, filtering the selection with a threshold, in order to detect and count the passage of vehicles along the road. We obtain a simple procedure, scalable, and adequately validated by the results for that purpose.

## INTRODUCTION

Knowing the number of vehicles in circulation is an interesting fact for the control of traffic and the statistical measures are useful for urban design planning and for the prevention and projection of actions appropriate to the traffic situation. To this end, the generality of municipalities and governments carry out vehicle counting programs to know and plan adequately actions depending on the needs that the volume and density of vehicle circulation requires.

Multiple techniques and methods have been developed and used for the detection of vehicles on highways and urban roads. They often require advanced computing techniques, costly instruments (magnetic loops [Oh 2001] or video cameras [Chellappa 2004]) or analysis that cannot be implemented simultaneously with the information acquisition procedure. It is even very common, especially in the urban area, to see people carrying out the arduous work of visually counting the road traffic by observation and direct annotation.

The detection of moving vehicles on busy roads through the analysis of the acoustic signal can be complicated due to the overlapping of the noise coming from the different vehicles. Even as in the case we present in this work, on single lane roads, it can be difficult to discern the presence several vehicles when they travel very close. Although the ear is usually able to distinguish the presence of more than one vehicle, by simply observing the acoustic signal this work is much more complicated. Sometimes what looks like a single vehicle are several, and vice versa.

In order to be able to carry out the detection and counting of the passage of vehicles, we developed a simple method for the detection of moving vehicles by direct analysis of the acoustic signal. This procedure allows an algorithmic low cost in situ detection. For this, we have performed an algorithm based on the evaluation of the negative minima of the second discrete derivative of the smoothed acoustic signal, filtered by a threshold; A technique widely used in the analysis of images for the detection of boundary contours, in which it is called the Gaussian laplacian (L.O.G.).

This study will be presented with the following framework: materials, methodology (background and hypothesis, proposed algorithm), results and discussion, and conclusions and future work.

## MATERIALS

We use a SQuadriga portable recorder from Head Acoustics® and its companion microphone. We place it on the shoulder, 1.5 meters from the edge of the road and 5 meters from the center of it. We recorded the traffic noise, placing the recorder at 1.75 meters high, with a sampling frequency of 48kHz, at 24bits. The road is single lane. Later the recorded signal is analyzed by own algorithms developed with the well-known software MatLab®. To contrast the results, we simultaneously placed a camcorder recording traffic on the road.

## METHODOLOGY

### Background and hypothesis

We assume that the maximum acoustic intensity by the passage of the vehicle will occur in front of the microphone (see figure 1), as observed in equation 1. In this work we also assume the reciprocal: in the maximums of the acoustic signal the information of the passage of the vehicles by the point of recording is found.

The acoustic intensity $I$ is proportional to the time average of the quadratic pressure $p^2$, and then:

$$I \propto \langle p^2 \rangle \propto \frac{1}{r^2} = \frac{1}{d^2 + v^2 t^2}, \qquad (1)$$

where it has been assumed that the vehicle velocity *v* is constant and that the origin of the time *t* is zero is just when the vehicle passes in front of the microphone.

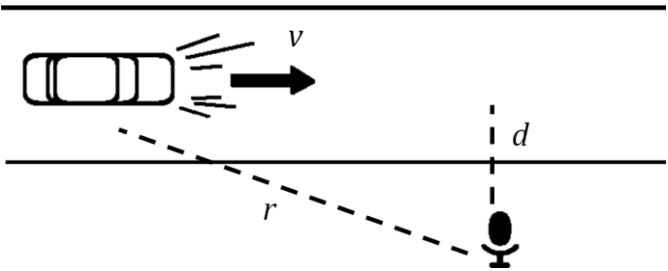

*Fig. 1: Measurement diagram. The relative situation of microphone, roadway and vehicle can be observed. d: = microphone distance to the center of the road, r: = distance from the microphone to the vehicle, v: = speed of the vehicle.*

By the described procedure of measurement we obtain a sound pressure signal. To automatically detect the passage of vehicles we perform a one-dimensional equivalent filtering procedure to the very extended laplacian of gaussian (L.O.G.) filtration in the treatment of two-dimensional images.

The procedure outlined in figure 2 is a simple and usual method for detecting borders in a signal. One can first observe: the elimination of noise by the convolution of a signal with a Gaussian; and second: border detection by the first derivative of the filtered signal.

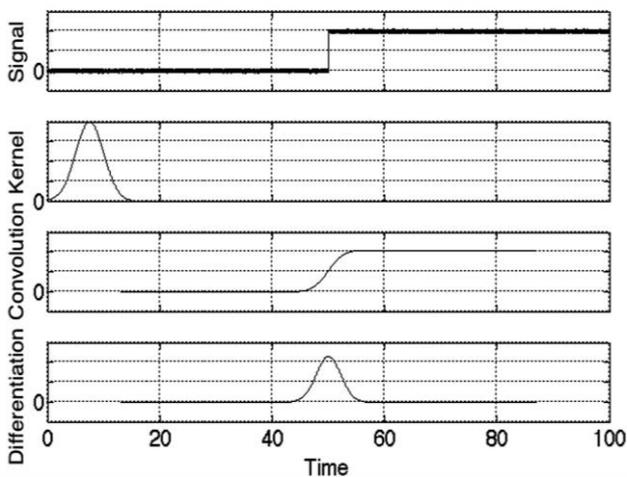

*Fig. 2: Procedure for the detection of borders in a 1-D signal. a) Original signal; b) Gaussian; c) Convolution of the original signal with the Gaussian (filtering); and d) Discrete temporal derivative of the convoluted signal (detection).*

While we want to detect the passage of a vehicle in front of the microphone, we also want to be able to discern the passage of several when this occurs, even though they pass relatively close to each other, and this information cannot be extracted from the detection of maxima after the first derivative Represented in the previous figure, since we would obtain a single maximum for the contribution of noise of the various cars that follow each other very closely.

However, while there may be a single maximum in the filtered acoustic signal, there will always be a certain delay of a vehicle with respect to the previous one, however small. We thus wish to extract information about the deformation that the acoustic signal suffers due to the superimposition of the non-simultaneous noise inputs of several cars. For this, we will not analyze the maximums in the filtered acoustic signal but the points of inflection or modulation in the slope. And we do this by analyzing the second discrete time derivative of the signal.

Due to the convolution of the original signal with the Gaussian, we assume in advance a limitation when the passage of two vehicles is almost simultaneous, which will make difficult the discernment of both, generating under-detection. To avoid this, the Gaussian must be narrowed down as much as possible, taking into account that the more it narrows, the less the signal is softened, so that sporadic noises and background noise confuse the measurements, generating over-detection.

And on the other hand, we also assume the possibility of obtaining more than one detection for a single vehicle; especially in cases of higher engine noise emitted, slower and more changeable driving (gear changes, large variations in acceleration, etc.).

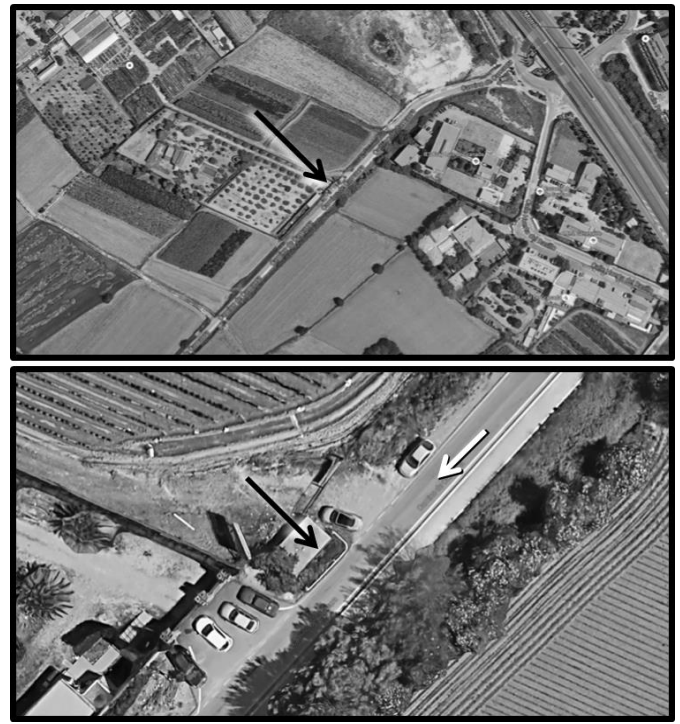

*Fig. 3: Measurement zone: Camino de Camaura, Granada (Andalusia, Spain). Coordinates: 37.175236 N, 3.616460 O. Images and coordinates obtained from Google Maps. Measured point marked with a black arrow. Direction of the vehicles indicated with a white arrow. Orientation: north-up, east-right.*

In order to face the first limitation in the discernment of several close vehicles, we set the microphone as close as possible to the road, so that the sound intensity of the cars increases with respect to background noise, and the separation between the sounds of two consecutive vehicles. As for the second limitation, we try to minimize it by choosing a relatively intermediate and non-curved zone along the roadway, looking for a more stable conduction (figure 3).

**Proposed Algorithm**

From the recording we get a *WAV* format file, with a sampling frequency $F_s = 48000 Hz$, corresponding to a vector $\boldsymbol{x} = (x_1, x_2, \ldots, x_p)$ of discrete sound pressure values, configuration defining a time vector $\boldsymbol{t} = \left(0, \frac{1}{F_s}, \frac{2}{F_s}, \ldots, \frac{p-1}{F_s}\right)$ for sampling, with $p$ a natural number that depends on the recording time.

We perform the detection in an algorithmic process whose steps are described below:

1.- We take the absolute value of the signal $\boldsymbol{x}$ $\boldsymbol{a_x} = (|x_1|, |x_2|, \ldots, |x_p|)$ to work with the amplitude of the acoustic pressure.

2.- We save the background noise information as a threshold. For this we select in $\boldsymbol{a_x}$ some sections in which there is no car passage, generating a vector of background noise $\boldsymbol{n} = (n_1, n_2, \ldots, n_l)$ with these sections, and we calculate the *average noise amplitude* $\bar{\boldsymbol{n}}$ as:

$$\bar{n} = \frac{1}{l} \sum_{i=1}^{l} n_i.$$

3.- We filter the vector $\boldsymbol{a_x}$, and obtain $\boldsymbol{w}$:

$$\boldsymbol{w} = \boldsymbol{a_x} * \boldsymbol{g},$$

by discrete convolution with a vector values of discrete normalized gaussian function $\boldsymbol{g} = \left(e^{\frac{-(t_i - t_c)^2}{2\sigma^2}}\right)_{i=1,2,\ldots,r}$, with $t_r = 2t_c$ (this is centered in $t_c$ and taking the first $r$ values of $\boldsymbol{t}$); resulting

$$w(k) = \sum_j a_x(j) g(k - j + 1),$$

with $j = M_1, M_1 + 1, \ldots, M_2$ ; where $M_1 = max(1, k + 1 - n)$, $M_2 = min(k, m)$, and where $m = dim(a_x)$ and $n = dim(g)$. The values of the index of $\boldsymbol{w}$ result in $k = 1, 2, \ldots, m + n - 1$.

4.- We calculate the second discrete numerical derivative of the filtered signal $\boldsymbol{w}$ and detect the negative minima of the second derivative (maxima of the filtered signal and inflection points or with modulation in the slope), according to:

$$w'(j) = \frac{w(j+1) - w(j)}{h_t},$$

$$w''(j) = \frac{w'(j+1) - w'(j)}{h_t},$$

this is by forward differences, with $h_t = {1}/{F_s}$.

5.- The detection points we select are those whose noise level exceed $q$ times the threshold of background noise $\bar{\boldsymbol{n}}$.

$q$-Value should be adjusted according to the characteristics of the medium, as well as the location of the microphone, as high as possible. If the acoustic signal of vehicles stands out very little above the background noise, the value $q$ should be adjusted close to the unit, reducing the likelihood of discrimination of vehicles with respect to noise. The more the noise is distinguished from the passage of vehicles with respect to background noise, the higher the value $q$ can be, and the detection of the vehicle passage will be better. Therefore it is convenient to analyze a section of the acoustic signal to ensure a good adjustment of the value $q$, adjusting it so as not to detect more or less vehicles of the recorded ones.

Thus we seek, first: by filtering the acoustic signal with a gaussian of sufficient width, we eliminate the influence of noise and sporadic noise sources; second: when performing the detection not in the filtered signal, but in the second derivative of the signal, we obtain information of those closed vehicles that in the filtering are confused, or by the own superposition of noise of several sources, analogous to how border regions are obtained in image analysis using L.O.G. filtering, and which is characterized by zones of modulation of the acoustic signal; and finally, by setting a sufficiently high threshold we should eliminate false events, such as the song of a bird, the sound of a horn, and so on.

**RESULTS AND DISCUSSION**

The procedure described above has been applied for a short recording interval of 89.5 seconds, adjusting the values of the model in order to obtain more precise results. The values of the constants we have introduced in the model are: $t_c = 3s$ and $q = 1.5$; both values high enough. It has then been applied for a total recording time of 21.5 minutes. It is important to note that the road under study is located 230 meters from a high-speed highway, in an area with much noise of birds and helicopters (during the recording the area was flown three times by a helicopter).

In figure 4 we show the filtering procedure (original signal, absolute value, signal filtered by convolution with gaussian, first derivative and second derivative of the filtered signal) of the original acoustic signal, applied to the 89.5 seconds test interval described above.

In the figure 5 we can see the results of the detection of vehicles applying the proposed procedure with the indicated variables to the test interval, and where it is represented first: the points detected as minimum of the second derivative of the filtered signal; second: the selected points (the definitive points: those points which exceed the established threshold); and third: the points over the acoustic signal. Of course, in this interval has been achieved the detection of all vehicles and none more.

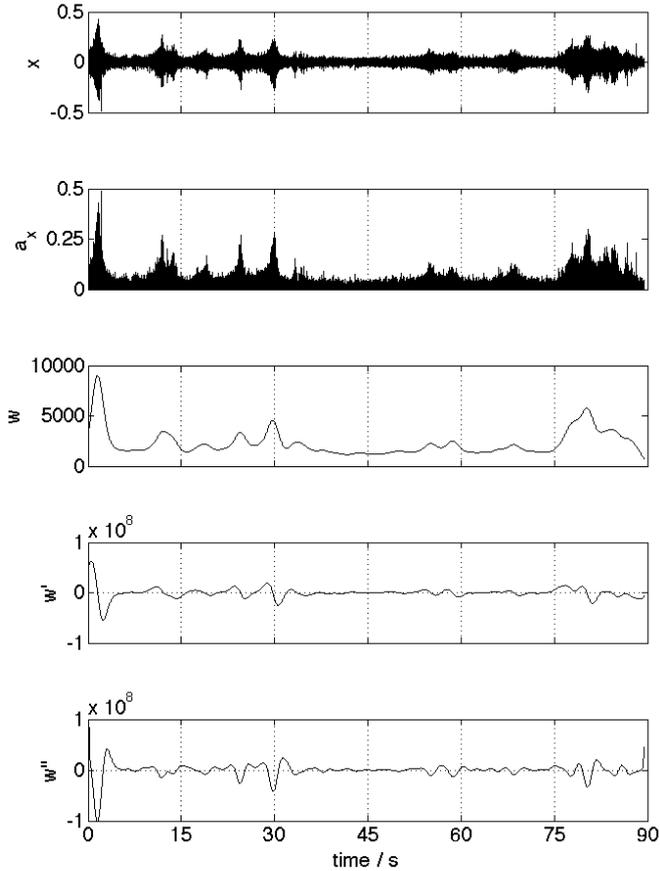

*Fig. 4: a) Acoustic signal; b) Absolute value of the acoustic signal; c) Absolute value filtered by convolution with gaussian; d) First derivative of the filtered signal; e) Second derivative of the filtered signal.*

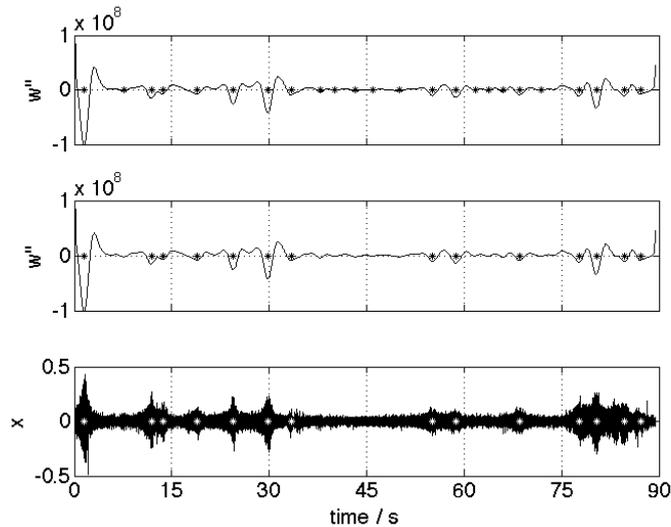

*Fig. 5: a) Second derivative of the filtered signal with '*' on negative minima; b) Second derivative of the filtered signal with '*' on the selected points (over the threshold); c) Original signal with '*' on the selected points.*

The results of the 21.5 minutes interval in which we have applied our procedure are shown in table 1, where we can see the number of vehicles *e* that have passed in front of the microphone (events), the number of vehicles detected *d* (detection), the number of vehicles that were computed but did not cross in front of the microphone *p* (false positives), the number of vehicles that were not computed *n* (false negatives) and the efficacy $\eta$ of the algorithm in the applied range (efficacy), defined as $\eta = d - p$. The table also shows the relative deviation from the total number of events, according to:

$$D.R. = \frac{a - e}{e}$$

where *a* is any of the measured parameters (*e, d, p* or *n*).

|  | Vehicles | R.D. |
|---|---|---|
| Events | 141 | 100% |
| Detection | 139 | 98.58% |
| False positives | 6 | 4.26% |
| False negatives | 8 | 5.67% |
| Efficacy | 133 | 94.33% |

*Tab. 1: Events (recorded vehicles), detection (detected vehicles), false positives, false negatives, efficacy; and relative deviation of them.*

**CONCLUSIONS AND FUTURE WORK**

We have developed an algorithmic procedure for the detection of passing vehicles, applied to the recording of the traffic noise of a single-lane roadway, using a simple methodology based on the two-dimensional L.O.G. filtering methodology, obtaining an efficiency of the 94.33% on detection over our measurement.

It is validated a procedure of great importance for being economical, easy to use and applicable not needing constructions or complex instrumentation but a microphone, present in many portable devices, which is useful for the studied case.

By the applied methodology the present development can be used to procedures of traffic maps modeling, as well as for the detection and counting of other events with acoustic mark.

As a future work the application and validation of the proposed algorithm to more complex roads (multi-lane, double-way) could be interesting and more useful.

Also it would be interesting to couple classification algorithms to vehicle detection, in order to characterize in detail the vehicular flow.